\documentclass[a4paper]{JHEP3}
\usepackage[centertags]{amsmath}
\usepackage{amssymb}
\usepackage{graphicx}

\title{A Study on Charged Neutron Star in $AdS_5$ }

\author{Vincenzo Parente$^{a}$\footnote{parente@na.infn.it}, Raju Roychowdhury$^{a,b}$\footnote{raju@na.infn.it}\\
\small{\emph{$^{a}$Dipartimento di Scienze Fisiche, Federico II University,\\
        Complesso Universitario di Monte S. Angelo,\\
        Via Cintia, Ed. 6, I-80126 Napoli, Italy\\
}}
\small{\emph{$^{b}$INFN - Sezione di Napoli, \\
        Complesso Universitario di Monte S. Angelo,\\
        Via Cintia, Ed. 6, I-80126 Napoli, Italy\\
}}
}

\abstract
{Motivated by an open question raised in recent times regarding the phase transition 
during  the collapse of a neutron star to form a black hole and related stability issues, 
we have constructed charged neutron stars in $AdS_5$ and show that these  stars 
become  unstable at a particular value of their radius, regarded as the Chandrasekhar radius. We 
reproduced the calculations recently done in \cite{Hartnoll:2010ref} in our $AdS_5$ charged star. 
The analysis shows that the non-Fermi liquid behavior found there in $AdS_4$ is still true in this 
higher dimensional case with the presence of Kosevich-Lifshitz oscillations.}
\keywords{} 
\preprint{DSF/NA/}

\begin{document}

\section{Introduction}
In recent times there has been a flurry of activity in the interplay between condensed matter physics  and AdS/CFT trying to bridge the gulf that exists in between them giving rise to several interesting
original ideas in this frontier (For an introduction to this field see for exmple \cite{Sachdev:2010ch,Hartnoll:2009qx,Hartnoll:2009sz,McGreevylect}). 
Recently in a series of papers \cite{JdB:2009, JdB:2010} the authors addressed the issue of a 
holographic description  of an astrophysical phenomena i.e. the collapse of a neutron star 
toward the formation of a  black hole beyond the so-called Oppenheimer-Volkoff (OV) limit 
\cite{OV:1939}. Near the very end 
of their article \cite{JdB:2009}, they have proposed an open question regarding the CFT meaning of the OV limit 
under consideration.The Neutron Star to Black Hole formation gives a hunch of a second-order 
quantum phase transition from a condensed matter physicist's view point turning a high density
baryonic state into a thermal QGP state and thus the issue needs proper investigation as 
suggested in \cite{JdB:2009, JdB:2010}. The main point of investigation is rather open ended and it 
would be really interesting to settle the issue whether a reliable realization of a degenerate 
fermionic star can indeed be achieved conclusively in the AdS/CFT duality framework. We have 
tried to formulate the very problem on our own way  rendering hope to give a partial, if not 
complete, explanation of the underlying phenomena which of course needs further study.

In this note we study charged neutron stars in $AdS_5$ space and focus on their 
collapse toward the formation of a black hole. This process is particularly 
interesting and worth studying from the holographic point of view (see \cite{JdB:2009, JdB:2010} and references therein).
This is because in terms of the boundary theory this process might as well symbolize a phase transition in $AdS$ space.

First we present a construction of a charged star in AdS space by adopting numerical methods.
Then we find that such a star ceases to exist at a particular value of the  mass, charge and radius. This fact implies that this star becomes unstable at this value of the mass and charge. This instability will trigger the formation of a new phase and (presumably) the star will collapse to form
a black hole. Now the behavior of a test fermion in the black hole background  has recently been analyzed in \cite{LMV:2009}. It is concluded from their analysis that  in this black hole phase the boundary theory has excitations about the Fermi-surface which is unlike the Fermi liquid.  
Further research on this point \cite{Larsen:2010} has shown that the Fermi vector depends exponentially on the scaling dimension if one uses the duality to extremal RN black holes while computing the non-Fermi liquid Green function. This proved the fact of violation of Luttinger theorem incase of non-Fermi liquids which is even true in case of the extremal $AdS_{5}$ RN Black Hole.
All such computations reported in \cite{LMV:2009, Science:2009, Hartnoll_strange, Lee:2009} revealed the presence of a Fermi surface in the bulk of a non-Fermi liquid type. For a much better 
and clear understanding of their results and the underlying phenomena that occurs, we thought 
that it might be worthwhile to explore the boundary description of a simpler system, that of an ideal Fermi gas in AdS without the presence of a Black Hole. \footnote{It is worth pointing out here that in \cite{Hartnoll:2010tavan} the authors evaluated the fermionic correlators on a background with fermions but without a Black Hole horizon.} 

We perform a similar kind of analysis by putting in a test fermion in the bulk with the charged neutron star serving as the background. From the dynamics of  such a bulk fermionic field one can compute the two point function of the boundary composite operator that corresponds to the bulk fermion. Although, we have not written down the boundary correlation function explicitly in the star background, but we gave a possible direction how to do that in practice. \footnote{The computation is difficult in the sense that the boundary theory under consideration is strongly interacting, and the neutron star to black hole collapse process is time-dependent and hence the radial holographic dimension can't be easily constructed on the CFT side.} 
We hope that, from the pole structure of these two point function, it will be evident that  in the hydrodynamic limit one gets either a Fermi liquid or a non-Fermi liquid type of behavior, nevertheless the explicit computation needs to be performed. Via the AdS/CFT dictionary for fermions \cite{Mueck:1998} the conformal dimension of the dual operator in the boundary CFT can be controlled by the bulk fermionic mass. In \cite{Science:2009}, it has been already found that by tuning the mass one can match the conformal dimension of the boundary operator to that of the free  fermion, and thus one obtains the spectral function exhibiting a peak which is consistent with a Fermi liquid behavior. It is quite clear from their analysis that one can deviate away from the Fermi liquid behavior by tuning the mass away from the free value as the mass was interpreted as the proxy for coupling in \cite{Science:2009}.

However, we recalled the fact that if one computes the fermionic correlation functions in charged $AdS_5$ black hole backgrounds using probe fermion and analyze the spectral peak as done in \cite{Larsen:2010}, one finds that the Fermi momentum depends exponentially on the probe fermion mass interpreted as a proxy for the coupling and thereby violates Luttinger theorem indicating a non-Fermi liquid behavior as was found in \cite{LMV:2009}. We then show, following \cite{Hartnoll:2010ref}, the presence of Kosevich-Lifshitz (KL) kind of oscillation in our $AdS_{5}$ charged neutron star setting. One must emphasize the fact that , we are working with ``global AdS''
i.e. the spatial sections are $S^{3}$ rather than planes which implies that the field theory under consideration is on $time \times S^{3}$ instead of being on Minkowski space. In \cite{Hartnoll:2009ref, Hartnoll:2010tavan, Hartnoll:2010ref} the authors 
studied electron stars working with a field theory defined on the Minkowski space whereas the 
authors of \cite{JdB:2009, JdB:2010} adopted our point of view.
\section{Construction of charged neutron stars}
\label{second}
In this section we present the construction of the charged neutron star in 
$AdS_5$. We will work with units in which the $AdS$ radius is set to unity.
We consider the following metric ansatz
\begin{equation}\label{metric}
 ds^2 = -A(r) dt^2 + B(r) dr^2 + r^2 d\omega_3^2.
\end{equation}
Besides the metric we have a dynamical gauge field ($H$) and a fermionic field ($\psi$) in this background.  We will work in the gauge where the radial component of the  gauge field is zero (i.e. $H_r = 0$). With this choice of gauge we write the gauge field as
\begin{equation}
 H(r) = h(r) dt.
\end{equation}
Further we consider fermionic species with mass $m$, charged under this
gauge field with charge $q$. However, we shall treat the fermions in a hydrodynamic
approximation. For this we will take the limit \footnote{before taking this 
limit we must also ensure that the central charge of the boundary CFT is taken 
to infinity (something similar to the large N limit for $\mathcal{N}=4$ Yang-Mills)
with all other quantities keeping fixed. This is required to ensure that 
all the multi-trace operators are suppressed. One way to think of this limit in
the bulk is to consider N species of fermions with the same mass and then take 
N to infinity keeping everything else fixed.} that the number of particles per 
AdS radius is infinite with $\epsilon_F/m$ fixed, $\epsilon_F$ being the Fermi energy.
The ideal degenerate Fermi gas formed by these particles can then be described
by the following hydrodynamic stress tensor for ideal fluids
\begin{equation}
\label{stresstensor}
 T_{\mu \nu} = (\rho +  p)u_{\mu} u_{\nu} + p g_{\mu \nu},
\end{equation}
where the pressure $p$ and the density $\rho$ are related to each other by the
flat space equation of state for the fermions i.e.
\begin{equation}
\label{EOS}
p + \rho = \mu n
\end{equation}
The use of the flat space  
approximation is again justified as there are large number of fermions within 
a given $AdS$ radius and therefore the fermions do not see the curvature of the space time.
In order to be consistent with this approximation, the mass and charge of the fermions must be
greater than the $AdS$ radius (i.e. unity in our case). This equation of state is implicitly 
given by
\begin{equation}\label{eqofstate}
p = \frac{(\mu - m)^3 \left(8 m^2+3 \mu  (3 m+\mu
   )\right)}{1920 \pi ^2},
\rho = \frac{2 m^5-5 m^2 \mu^3 + 3 \mu^5}{480 \pi ^2},
\end{equation}
where $\mu$ is the chemical potential for the fermions, which can be expressed in 
terms of the Fermi momentum ($k_F$) by the relation
$$ \mu = \sqrt{k_F^2 + m^2}.$$
In this flat space approximation for the fermions the particle number density ($\tilde{n}$)
is given in terms of the volume of the Fermi surface and we have
\begin{equation}
 \tilde{n} = \frac{\left(\mu^2 - m^2\right)^2}{128 \pi ^2}.
\end{equation}
Then it immediately follows that the charge density ($n$) is given by
\begin{equation}
 n = q \tilde{n} = q \frac{\left(\mu^2 - m^2\right)^2}{128 \pi ^2}.
\end{equation}
Note that in our problem these relations are locally valid and generally 
$\mu, n, p$ and $\rho$ are functions of $r$-coordinate \footnote{the symmetries 
of AdS prevent these quantities from becoming functions of the other coordinates} 
which we ultimately solve for. Here we would like to draw attention of the reader to 
a subtle point. $u_{\mu}$ as defined in (\ref{stresstensor}) is a static velocity field: 
$u_{\mu}dx^{\mu} = A(r)dt$ and the radial profiles of $p$ and $\rho$ are determined
by imposing the stress energy conservation which leads to the following condition:
\begin{equation}
\label{cond}
\frac{dp}{dr} + \frac{1}{A}\frac{dA}{dr}(\rho + p) = 0
\end{equation} 
This equation is very easy to solve. By making use of (\ref{EOS}) one easily verifies 
the fact that (\ref{cond}) is satisfied while the chemical potential obeys
\begin{equation}
\mu(r) = \frac{\epsilon_{F}}{A(r)}
\end{equation}
where at this stage $\epsilon_{F}$ is an arbitrary constant. Thus the radial dependence of 
the chemical potential $\mu$ is simply due to the gravitational redshift.
\subsection{The Equations to be solved} 
In the above set up we now write down the dynamical equations which we must 
solve in order to obtain the Neutron star numerically.

Firstly we have the Einstein equations. The two non-trivial Einstein equations are
obtained from the rr-component and the tt-component and they are respectively given by
\begin{equation}\label{eineqs}
\begin{split}
& \quad \quad 3 r \left(2 A'(r)+r h'(r)^2\right)-4 A(r) \left(B(r) \left(r^2
   p(r)+6 r^2+3\right)-3\right) = 0, \\
& A(r) \left(6 r B'(r)-4 B(r)^2 \left(r^2 \rho (r)-6
   r^2-3\right)-12 B(r)\right)-3 r^2 B(r) h'(r)^2 = 0,
\end{split}
\end{equation}
where as mentioned before we consider the pressure and the density as functions
of the radial coordinate.

Then we consider the Maxwell equations. In this case the non-trivial equation stems out 
from the t-component (which is a mere generalization of Coulombs law). This equation
is given by
\begin{equation}\label{maxeq}
 \left(\frac{A'(r)}{A(r)}-\frac{6}{r}\right) h'(r)+\frac{B'(r)
   h'(r)}{B(r)}-2 B(r) n(r)-2 h''(r) = 0.
\end{equation}

Finally we have to consider the equation of motion for the 
fermions. However, since the fermions are treated in a hydrodynamic
approximation this equation is the conservation of the stress tensor\footnote{
which is the relativistic version of the Navier-Stokes equation.}.
In this case the radial component yields the non-trivial equation
and is given by
\begin{equation}\label{hydroeq}
\begin{split}
 & A(r) \left(2 r B(r)^2 A'(r) (p(r)+\rho (r))  +3 r B'(r) h'(r)^2+ \right. \\ &\left. -6
   B(r) h'(r) \left(r h''(r)+3 h'(r)\right)\right) +3 r B(r)
   A'(r) h'(r)^2+4 r A(r)^2 B(r)^2 p'(r) = 0.
\end{split}
\end{equation}

In these equations both $p(r)$ as well as $\rho(r)$ are present explicitly. However, we will
eliminate both $p(r)$ and $\rho(r)$ in terms of $\mu(r)$ with the help of the equation of 
state \eqref{eqofstate}. Now these equations are extremely non-linear and we have to resort 
to numerical means in order to solve them.
\subsection{Solving the equations}
One obvious solution to these equations is the charged black hole 
in $AdS_5$ with the chemical potential being constant throughout the
space. In terms of the above mentioned functions this solution 
may be written as \cite{Emparan:1999},
\begin{equation}\label{blackholesol}
 \begin{split}
  A(r)&= \left( 1+r^2\left( 1-\frac{M}{r^4} +\frac{Q^2}{r^6} \right) \right), \\
  B(r)&= \left( 1+r^2\left( 1-\frac{M}{r^4} +\frac{Q^2}{r^6} \right) \right)^{-1}, \\
  \mu(r) &= m, \\
  h(r) &= \mu_B - \frac{Q}{r^2}.
 \end{split}
\end{equation}
where $m$ and $\mu_B$ are constants. The parameter $m$ is related to the ADM 
mass of the Hole as $M = \frac{3 \omega_3}{16 \pi G}m$ \cite{Emparan:1999} where in natural 
units $16 \pi G=1$ and $\omega_3$ is the volume of the 3-sphere. Also, $\mu_B$ is the 
electrostatic potential difference between the horizon and infinity. Since the constant 
value of the chemical potential outside the star is $m$, the mass of the fermionic species, 
therefore we use the same notation here also.
Similarly we use $\mu_B$ to denote the constant part of the gauge field
as it ultimately turns out to be the boundary chemical potential.

Now it is expected that outside the neutron star our solutions should reduce to
the black hole solution. Therefore, we shall obtain a solution inside the neutron star 
and then patch up our solution with this black hole solution outside. Thus we start with
a boundary condition at the origin (which may be thought of as the centre of the star)
and make a choice of the time coordinate inside the star such that at the boundary the first
derivative of the field strengths match. In order to determine the consistent boundary 
conditions at the origin we solve the equations about $r=0$ and find that,

\begin{equation}\label{solnear0}
\begin{split}
 A(r) &= A_0 + \frac{1}{180} r^2 \left(-\frac{A_0 m^5}{8 \pi ^2} +\frac{15
   A_0 m^4 \mu_0}{32 \pi ^2}-\right. \\ & \qquad \qquad \qquad \qquad \qquad \qquad\left. \frac{5 A_0
   m^2 \mu_0^3}{8 \pi ^2}+\frac{9 A_0 \mu_0^5}{32 \pi ^2}+180 A_0\right) + O(r^3),\\
 B(r) &= 1 + \frac{1}{90} r^2 \left(\frac{m^5}{16 \pi ^2}-\frac{5 m^2
   \mu_0^3}{32 \pi ^2}+\frac{3 \mu_0^5}{32 \pi
   ^2}-90\right)+ O(r^3),\\
 \mu(r) &=\mu_0+\frac{r^2}{92160 \pi ^2 A_0} (32 A_0 m^5 \mu_0-120 A_0
   m^4 \mu_0^2+160 A_0 m^2 \mu_0^4-72
   A_0 \mu_0^6  \\ 
    & -46080 \pi ^2 A_0 \mu_0 +135 m^4 q^2-270 m^2 q^2 \mu_0^2+135 q^2
   \mu_0^4 ) + O(r^3),\\
 h(r) &=\mu_B -\frac{q r^2 \left(m^2-\mu_0^2\right)^2}{1024 \pi ^2}+ O(r^3),
\end{split}
\end{equation}
solve the equations \eqref{eineqs},\eqref{maxeq} and \eqref{hydroeq} upto $O(r^2)$.
Here $A_0$ and $\mu_0$ are the values of $A(r)$ and $\mu(r)$ at the origin.
These are the parameters of our problem and we have to choose values for these
parameters which serve as initial values of our differential equations.
Again $m$ and $q$ are the mass and charge of a single species of fermion. 
The most striking thing to note about this solution is that the value of 
$B(r)$ at the origin is fixed to be unity. We do not have the freedom to 
choose this value on independent grounds. Also the parameter $q$ in (\ref{solnear0})
yield the charge of the black hole as $Q = \frac {2\sqrt{3}\omega_3}{8 \pi G} q$ \cite{Emparan:1999}.
\subsubsection{Numerical Solution of the equations}
We proceed to solve these equations numerically in the following way.
At first we fix the value of the chemical potential at the origin to be $\mu_0$.
Then we fix the value of $A_0$ to be unity and then fix a scale for the time 
coordinate in the patch inside the star so as to meet the boundary conditions
$A(r)=1/B(r)$ at the radius of the star ($R$). The radius of the star is obtained 
from the value of $r$ where the density $\rho(r)$ goes to zero or the 
chemical potential $\mu(r)$ goes to $m$ as can be easily checked from the 
equations (\ref{eqofstate}) and (\ref{blackholesol}). Further from the value of 
$A(R)$ (which is the same as $1/B(R)$), together with the matching condition for the
Electric field at $r=R$, we determine the mass and charge of the black hole solution
with which we patch up outside the neutron star. Note that in this procedure, matching of 
the first derivative of $A(r)$ (which is the gravitational field strength in a 
rough sense) at $r=R$ is automatic. As a part of our choice of units we
take the $AdS$ radius to be unity.

\begin{figure}
 \begin{center}
  \includegraphics[width=\textwidth]{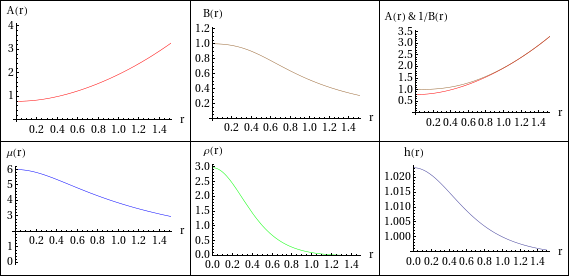}
 \end{center}
\caption{Solution of various functions inside the star. The values of the 
parameters are $m=3$, $q=1$ and $\mu_0=6$. We have shown the plots upto r=1.48 
which is the critical radius of the star.}
\label{fig:starsol}
\end{figure}

In fig:\ref{fig:starsol}, we present the solution when the parameters $m=3$, $q=1$ and $\mu_0=6$. 
This corresponds to the core density of 2.97. The Mass of the star is 0.073 and the 
charge of the star is 0.0083 \footnote{These values of the charge and mass 
should not be directly compared with the mass and charge of a single fermion
since in this case we should also consider the Newton's constant which we 
have not included in the present analysis.}.The radius of the star is 1.48.
Note that the density function $\rho(r)$ goes to zero at the edge of the star.
Also the values of $A(r)$ and $1/B(r)$ match at the end of the star.

\subsubsection{A class of Neutron stars}

\begin{figure}
\begin{center}
\includegraphics[width=0.5\textwidth]{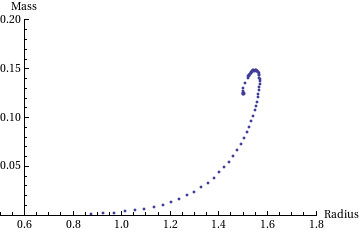}
\end{center}
\caption{Plot for Mass vs the radius of the neutron star as the core density is varied. Here also 
we have taken $m=3$ and $q=1$.}
\label{fig:massVSrad}
\end{figure}

Now we can vary the value of the chemical potential at the origin to obtain a 
large class of neutron stars. This amounts to obtaining neutron stars for various 
values of core density. Further under a variation of the core density (which varies
form $0$ to $\infty$) we can make a plot of the mass and radius of the star (see fig:\ref{fig:massVSrad}).
In this plot we find there exists a maximum value of mass for the star sometimes termed as critical mass of the star. \footnote{We can render a holographic interpretation to this critical mass, too. As found in \cite{JdB:2010} the limiting mass in the boundary CFT theory translates to a limiting conformal dimension of the composite operator made out of the fermionic primary fields which in the large N limit construct our model degenerate star.}
Further the solution ceases to exist at a particular value of the mass and radius. This is the
signature of a critical behavior and occurs when the density at the origin 
approaches infinity. At this critical point the neutron star is expected to start 
collapsing into a black hole.

\begin{figure}
\begin{center}
\includegraphics[width=0.5\textwidth]{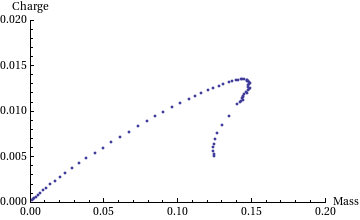}
\end{center}
\caption{Plot for charge vs mass of the neutron star as the core density is varied. Here also 
we have taken $m=3$ and $q=1$}
\label{fig:chargeVSmass}
\end{figure}

In a similar way, we can obtain a plot for the mass vs charge of the star
(see fig:\ref{fig:chargeVSmass}). In this plot we see that for a given value of charge two solutions 
exist corresponding to two different masses. Presumably the one with more mass is the 
stable one, the other being unstable.

We again remind the readers of the intriguing fact that working in ``global AdS'' we have $S^{3}$ as 
spatial section thus providing a scale in the field theory, which is the radius say $R$ of $S^{3}$. 
There is yet another scale in the problem, the chemical potential $\mu$. The related phase 
transition from the star to a black hole occurs due to the competition between the two scales. In 
particular at the critical point of phase transition one would get $R_{S^{3}} \sim \frac{1}{\mu}$.
So the very existence of $R_{S^{3}}$ is crucial for our purpose.

In recent years, all the various works \cite{LMV:2009, Larsen:2010, Hartnoll:2009ref, Hartnoll:2010ref} on Fermi surfaces consider Minkowski space while 
considering the boundary field theory. This was important in order to have a well-defined 
momentum $k$. On a sphere $k$ is not a good quantum
number, in particular if $R_{S^{3}} \sim \frac{1}{\mu}$ one might worry about whether the notion
of a Fermi surface makes sense in the momentum space. The answer to this is, yes, indeed the
notion can be justified. As we mentioned in Sec.\ref{second}, that in our choice of units the
charge and mass of the fermion are large, so that the boundary field theory fermions are locally
in a flat space making sense of a Fermi momentum in approximate terms.
\section{Holographic dual of the charged star}
We now study the holographic description of this neutron star solution.
By studying the dynamics of a probe fermion in this background one could get a hint at 
the boundary description of this charged neutron star. 

%
\subsection{Dirac Equation in Spherically symmetric space time}
We first present the Dirac equations in spherically symmetric space times
which we subsequently solve in our star background \cite{Ortega:2009}. \footnote{One should pay attention to the fact that while the fermions are free in the bulk, they are not so on the boundary (i.e. they don't obey Dirac equation on the boundary) as the boundary theory is strongly coupled.}

We have to incorporate the 
interaction of the probe fermion with the fermions forming the star. This we 
accomplish by considering a new gauge (different from the existing one)
with only a non-zero time component being the chemical potential,
due to the fermions forming the star. We justify this procedure in Appendix \ref{ChemPot}.

We consider the metric to be of the form (as in \eqref{metric})
\begin{equation}\label{metricnew}
 ds^2 = -A(r) dt^2 + B(r) dr^2 + r^2 \left(d\theta^2+\sin^2{\theta} d\phi^2+
            \sin^2{\phi} \sin^2{\theta} d\lambda^2 \right),
\end{equation}
The natural choice of vielbein basis vectors are
\begin{equation}
 e^{\tilde{t}}_{t}= \sqrt{A};\quad e^{\tilde{r}}_{r}= \sqrt{B}; \quad
 e^{\tilde{\theta}}_{\theta}= r; \quad e^{\tilde{\phi}}_{\phi}=r \sin {\theta};
 \quad e^{\tilde{\lambda}}_{\lambda} = r \sin{\theta} \sin{\phi}.
\end{equation}
where $\{\tilde{t},\tilde{r},\tilde{\phi},\tilde{\theta},\tilde{\lambda}\}$ are 
the tangent space coordinates.
Then the non zero components of the spin connection are
\begin{equation}
 \begin{split}
  \omega_{t}^{\tilde{t},\tilde{r}} &= -\frac{1}{2} \frac{A'(r)}{\sqrt{A ~ B}};\quad
  \omega_{\theta}^{\tilde{r},\tilde{\theta}} = \frac{1}{\sqrt{B}}; \\
  \omega_{\lambda}^{\tilde{r},\tilde{\lambda}} &= \frac{\sin{\theta} \sin{\phi}}{\sqrt{B}};\quad
  \omega_{\phi}^{\tilde{r},\tilde{\phi}} = \frac{\sin{\theta}}{\sqrt{B}}; \\
  \omega_{\lambda}^{\tilde{\theta},\tilde{\lambda}} = \cos{\theta} & \sin{\phi};\quad
  \omega_{\lambda}^{\tilde{\phi},\tilde{\lambda}} = \cos{\phi};\quad
  \omega_{\phi}^{\tilde{\theta},\tilde{\phi}} = \cos{\theta}.
 \end{split}
\end{equation}
Using the above equations the Dirac equation $$(\displaystyle{\not}D -m) \psi = 0, $$
reduces to
\begin{equation}\label{direq}
\left(\Gamma^{\tilde{r}} \frac{1}{\sqrt{B}}\partial_r + \Gamma^{\tilde{t}} \frac{1}{\sqrt{A}} \nabla_t
 - \frac{1}{2} \frac{1}{\sqrt{B}} \left( \frac{A'(r)}{2 A} +\frac{3}{r}\right) \Gamma^{\tilde{r}} 
 +\frac{1}{r} \displaystyle{\not}D^{(S^3)} -m \right) \psi = 0.
\end{equation}
where $\displaystyle{\not}D^{(S^3)}$ is the Dirac operator on the three sphere. The eigenvalues of this operator are \cite{Camporesi:1995}
\begin{equation}
 \kappa = \pm i \left( \frac{3}{2} + n \right), \quad n \in \{0,1,2, \dots\}.
\end{equation}
We shall denote the eigenfunction corresponding to these eigenvalues by $f_{\kappa}$.
Also the $\nabla_t$ operator denote the gauge covariant derivative and contains both the gauge
field and its time derivative. All the gamma matrices in \eqref{direq} have a tangent 
space index and hence are flat space gamma matrices.
Here we use the following basis for the gamma matrices.
\begin{eqnarray}
\Gamma^{\tilde{r}} = \mathbb{I} \otimes (-\sigma_3); \quad
\Gamma^{\tilde{t}} = \mathbb{I} \otimes (i \sigma_1); \quad
\Gamma^{\tilde{i}} = \sigma^{i} \otimes \sigma_2.
\end{eqnarray}
where $\sigma^{i}$ are the ordinary Pauli matrices.
Further we consider the following separation of variables
\begin{equation}
 \psi = f_{\kappa} \left( \begin{array}{c} \phi_{+} \\ \phi_{-} \end{array}\right) e^{-i \omega t}.
\end{equation}
where $\phi_{\pm}$ are the eigenvectors of $\Gamma^{\tilde{r}}$ with 
eigenvalues $\pm 1$. Using this separation of variables the Dirac equation
\eqref{direq} reduces to the following set of coupled first order equations
\begin{equation}\label{direqsimp}
 \begin{split}
 & \left( \partial_r + \frac{1}{2} \left( \frac{A'(r)}{2 \sqrt{A}} + \frac{3}{r}\right)
     + m \sqrt{B} \right) \phi_+ - \sqrt{B} \left( \frac{u(r)}{\sqrt{A}}-\frac{k}{r}\right) \phi_- =0.\\
 & \left( \partial_r + \frac{1}{2} \left( \frac{A'(r)}{2 \sqrt{A}} + \frac{3}{r}\right)
     - m \sqrt{B} \right) \phi_- + \sqrt{B} \left( \frac{u(r)}{\sqrt{A}}+\frac{k}{r}\right) \phi_+ =0.
 \end{split}
\end{equation}
where $k = i \kappa$, and $u(r)=\omega + \mu(r) - 3/2 q h(r)$ \footnote{Note that the 
unconventional factor of $3/2$ is present because here we are using a different normalization of charge.}.
\subsection{The flow equation}
In this subsection we will derive and analyze the so-called flow equation for the Dirac fermions in $AdS_{5}$. Using the set of equations \ref{direqsimp}, one can define the wave function at small $r$ to be of the form:
(Note that $m$ is the fermionic mass, and it enters in the expansion)  
\begin{equation}\label{solform}
\begin{split}
 \phi_- &= \alpha \left( r^{mL}+\dots \right) + \beta \left( r^{-(mL+1)} + \dots \right),\\
 \phi_+ &= \gamma \left( r^{-mL}+\dots \right) + \delta \left( r^{(mL-1)} + \dots \right),\\
\end{split}
\end{equation}
Here $L$ is the $AdS$ radius that we have taken to be unity $L=1$. 
The coefficients $\alpha$, $\beta$, $\gamma$ and $\delta$ are what the authors of \cite{Larsen:2010} call A, B, C, D in their article and they are related with one another.
The retarded Green function could be written as \cite{LMV:2009, Larsen:2010}
\begin{equation}
\label{green}
G_R=\epsilon^{-2m}\lim_{\substack{\epsilon\rightarrow0}}
\left . \begin{pmatrix}
\xi_+&0\\
0&\xi_-
\end{pmatrix}\right|_{1/\epsilon}
\end{equation}
where $\xi_-$ is defined as suitably defined ratio between $\phi_-$ and $\phi_+$.
\begin{equation}
\xi_-=-i\frac{\phi_-}{\phi_+}
\end{equation}
From the system of equations, \eqref{direqsimp} the flow equation can be derived, dividing the second equation by $\phi_+$ and inserting into it the first one, finally yielding:
\begin{equation}
\label{starflow}
i\partial_r\xi_--2i\xi_-m\sqrt{B}-\xi_-^2\sqrt{B} \left( \frac{u(r)}{\sqrt{A}}-\frac{k}{r}\right)+\sqrt{B} \left( \frac{u(r)}{\sqrt{A}}+\frac{k}{r}\right)=0
\end{equation}
The numerical solution of this equation has some singularity problems in $r$, due to the choice of boundary conditions, thus preventing its use in the calculation of the correlator in the neutron star background. This issue needs further careful investigation.

\subsubsection{The initial conditions}\label{bdycond}
Now we have to specify the boundary conditions for the equations \eqref{direqsimp}.
This is done by demanding regularity of the solution near the origin ($r=0$).
This regularity criterion in general depends on the value of $k$ once we fix the 
mass and charge of the fermion. The lowest positive value is $k=3/2$, in which we will
focus for our present purpose. In this case the regularity at the origin demands,
that if $\phi_+$ is $1$ at  the origin then $\phi_-$ should be $-1$ i.e.
$$ \phi_{+}(r=0) = 1\;\;\,;\;\;  \phi_{-}(r=0) = -1$$
We shall use this boundary condition to solve the equations \eqref{direqsimp}.
\begin{figure}
 \begin{center}
  \includegraphics[width=0.7\textwidth]{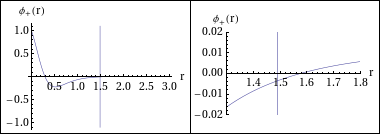}
 \end{center}
\caption{Plot of $\phi_+$ in the star background constructed in the previous section.
The vertical line denotes the value of $r$  for the edge of the star.
Here the values of the parameters are $m=3$, $q=1$, $\mu_B=1$, $k=3/2$, and $\omega=3$.
In the  right image we focus on the edge where the first derivative of the function is continuous.}
\label{fig:phip}
\end{figure}
\subsubsection{Numerical solution of the Dirac equation}\label{thesol}
\begin{figure}
 \begin{center}
  \includegraphics[width=0.7\textwidth]{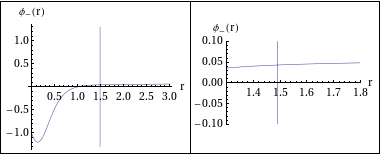}
 \end{center}
\caption{Plot of $\phi_-$ in the star background constructed in the previous section.
Here the values of the parameters are also $m=3$, $q=1$, $\mu_B=1$, $k=3/2$, and $\omega=3$. 
Again the first derivative of $\phi_-$ is continuous as can be seen from the plot on the left.}
\label{fig:phim}
\end{figure}

The solution of the probe fermion in the neutron star background is obtained numerically
for $m=3$, $q=1$ and $k=3/2$. It is shown in figs:\ref{fig:phip} \& \ref{fig:phim}. Here we use the 
boundary conditions as discussed in \S\S\ref{bdycond}. We at first let the differential equation 
evolve to obtain a solution up to the end of the star and subsequently use the value 
of $\phi_{\pm}$ at this edge as the initial value for the subsequent evolution 
outside the star. The first derivative match of the solution in these two patches
is automatic (see figs:\ref{fig:phip} \& \ref{fig:phim}) and directly follows from the fact that all the 
 functions that appear in the equations \eqref{direqsimp} are continuous at  the edge of the star.
%
One possible direction, one can take from here, is that, one might try to compute the boundary correlation function in the presence of the star and understand the behavior of the bulk Fermi surface in terms of the boundary correlators, taking the same route outlined in the formalism of \cite{Iqbal:2009}. A noteworthy point here, is that the only difference, we should care about, is that we need to evaluate the fermionic correlators in the neutron star background, whereas there exist large number of literature doing the same computation in the background of a charged black hole (See for example \cite{LMV:2009,Science:2009,Larsen:2010,Hartnoll_strange,Lee:2009}.) 
\subsection{Kosevich-Lifshitz oscillations in $AdS_5$ neutron star}
It has been shown recently in \cite{Hartnoll:2010ref} that, the charged star shares some important features with the Fermi liquid. In particular, the presence of the Kosevich-Lifshitz oscillations \cite{Hartnoll:2009ref} in magnetic field has been shown in the case of AdS$_4$ charged fermion star. As said in \cite{Hartnoll:2010ref} the arguments used to demonstrate this feature could be generalized to higher dimensions, showing that even in AdS$_5$ Kosevich-Lifshitz oscillations are present, nevertheless the Luttinger theorem is violated. In order to calculate the magnetic susceptibility $\chi=-\frac{\partial^2\Omega}{\partial B^2}$ we can use the relation between the free energy and the total charge in the dual CFT as in \cite{Hartnoll:2010tavan, Hartnoll:2010ref}:
\begin{equation}\label{freeenergy}
\hat{\Omega}\propto\hat{Q}= \int_0^{r_s}\sqrt{g(s)}r^3\sigma(r)dr\;\;
\end{equation}
where in general the charge density is
\begin{equation}
\sigma=\int_{m}^\mu g(E)\nu(E)dE
\end{equation}
and the density of states in the full theory is
\begin{equation}
g(E)=\beta E(E^2-m^2-\ell B)
\end{equation}
The easiest way to generalize the argument as presented in \cite{Hartnoll:2010ref} to the higher dimensional case is to add an extra dimension, i.e. by considering the cylindrical-like metric
\begin{equation}
ds^2=-fdt^2+\frac{1}{f}dr^2+r^2d\omega_2+dz^2
\end{equation}
and then the integral expressing the charge density is generalized in $AdS_{5}$ just by adding an integration over $p_z$
\begin{equation}
\sigma=\frac{\beta B}{2}\int_{-\infty}^{\infty} dp_z\sum_l\int_{-\infty}^{\infty} dp\sum_l\frac{1}{1+e^{\frac{E(l,p,p_z)-\mu}{T}}}
\end{equation}
The energy, as in the $AdS_{4}$ case, is
\begin{equation}
E(l,p,p_z)=\sqrt{p^2+p_z^2+\ell B_{loc}}
\end{equation}
The local physical quantities are related to the dual field theory quantities as \cite{Hartnoll:2010ref}:
\begin{equation}
\begin{split}
B_{loc}&\propto\frac{1}{r^2}\hat{B}\\
T_{loc}&=\frac{\hat{T}}{\sqrt{g_{tt}}}
\end{split}
\end{equation}
In the following calculation we will drop out this subscript, and restore the correct dependence on physical quantities back to position at a later point. The Poisson resummation over Landau levels gives
\begin{equation}\label{sigma}
\sigma=\frac{\beta B}{2}\sum_k\int_0^\infty dl\int_{-\infty}^{\infty} dp_z\int_{-\infty}^{\infty} dp\sum_l\frac{1}{1+e^{\frac{E(l,p,p_z)-\mu}{T}}}
\end{equation}
Now expanding the Fermi distribution in terms of Matsubara frequencies and transforming the integral over $\ell$ into an integral over energies the integral (\ref{sigma}) becomes
\begin{equation}
\begin{split}
\sigma&=-\frac{\beta T}{2}\sum_{k,n}\int_{-\infty}^{\infty} dp_z\int_{-\infty}^{\infty} dp\:e^{-\pi i k (p^2+p_z^2+m^2)/B}\int_{\sqrt{p^2+p_z^2+m^2}}^\infty dE E \frac{e^{\pi i k E^2/B}}{E-\mu(1+i\frac{T}{\mu}2\pi(n+\frac{1}{2}))}\\
\end{split}
\end{equation}
This integral could be performed rotating the integration path so that the exponential in the numerator becomes real and decreasing, i.e. considering the rotation of the energy as
$E\rightarrow E e^{i\pi/4}$ if $k>0$ or $E\rightarrow E e^{-i\pi/4}$ if $k<0$, and remembering that the lower integration limit is set to $\sqrt{p^2+p_z^2+m^2}$. In this analytical continuation one has to take into account the role of the poles given by Matsubara frequencies. These poles contribute only if $p^2+p_z^2\leq\mu^2-m^2$ and in addition $n+1/2>0$ if $k>0$ and $n+1/2<0$ if $k<0$. The contribution of the poles is the one giving the oscillatory part of the charge density $\sigma_{osc}$, and thus it' s the only one we will consider next, hence through the Residue theorem then the $\sigma_{osc}$ is
\begin{equation}
\begin{split}
\sigma_{osc}&=2\pi\beta T\mu \:\:\text{Im}\sum_{k,n>0}e^{i\pi k (\mu^2-m^2)/B}e^{-4\pi^2k\left(n-\frac{1}{2}\right)T\mu/B}\iint_{D} dp_z dp\: e^{-i\pi k (p^2+p_z^2)/B}
\end{split}
\end{equation}
while the last integration is made in the region $D$, being the circle of radius $\sqrt{\mu^2-m^2}$ in the plane $(p,p_z)$. This expression could be inserted in \eqref{freeenergy}, thus obtaining the oscillating part of the free energy
\begin{equation}
\begin{split}
\hat{\Omega}_{osc}\propto\hat{Q}_{osc}=2\pi\beta T\mu\:\:\text{Im}\sum_{k,n>0}\int_0^{r_S} dr r^3\sqrt{g(r)}& e^{i\pi k (\mu^2-m^2)/B}e^{-4\pi^2k\left(n-\frac{1}{2}\right)T\mu/B}\times\\
&\times\iint_{D} dp_z dp\: e^{-i\pi k (p^2+p_z^2)/B}
\end{split}
\end{equation}
Since $\mu\gg B$, a saddle point approximation could be done. The $r$ dependence of the fast oscillating exponential is
\begin{equation}
e^{i\pi k (\mu^2-m^2)/B}=\exp i\pi k \left(\frac{h^2}{f}-m^2\right)\frac{r^2}{\hat{B}}=\exp{ik\frac{A(r)}{\hat{B}}}
\end{equation}
With this approximation the integral (\ref{freeenergy}) becomes
\begin{equation}\label{chargeosc}
\begin{split}
\hat{Q}_{osc}=2\pi\beta\:\:\text{Im}\sum_{k,n>0}r^{*}\sqrt{g(r^*)}\frac{m^*e^{i\pi/4}\sqrt{\hat{B}}}{\sqrt{A''(r^*)}}&e^{i\pi k\frac{A(r^*)}{\hat{B}}}e^{-4\pi^2k\left(n-\frac{1}{2}\right)m^*T/B}\times\\
&\times\iint_{D} dp_z dp\: e^{-i\pi k (p^2+p_z^2)/B}
\end{split}
\end{equation}
where, as usual, $r^*$ is the value of the radius where the argument of the exponential has an extremum. The last integration could be done passing to circular coordinates in the $(p,p_z)$ plane inserting the radial coordinate $\rho=\sqrt{p^2+p_z^2}$ as
\begin{equation}
\iint_{D} dp_z dp\: e^{-i\pi k (p^2+p_z^2)/B}=\int_0^{b} \rho d\rho e^{-i\pi k \rho^2/B}=e^{-i\frac{b^2\pi}{2B}}\frac{\sin\frac{\pi b^2}{2B}}{\pi}
\end{equation}
where $b=\sqrt{\mu^2-m^2}$ has been defined. The oscillatory part of the total charge is then obtained introducing this result in \eqref{chargeosc}. The summation over the index $n$ could be done as usual, giving an hyperbolic sine, moreover the sum over $k$ could be suppressed considering only the term $k=1$, since the other terms are negligible. In the end after all dusts get settled the oscillating free energy is
\begin{equation}
\begin{split}
\hat{\Omega}\propto \hat{Q}_{osc}&=2\pi\beta\:\:\text{Im}\left[\frac{1}{r^{*}}\sqrt{g(r^*)}\frac{m^*\hat{B}^{3/2}}{\sqrt{A''(r^*)}}e^{i\pi \frac{A(r^*)}{\hat{B}}}e^{-i\frac{b^2\pi}{2B}}\frac{\sin\frac{\pi b^2}{2B}}{\pi}\sum_{n>0}e^{-4\pi^2\left(n-\frac{1}{2}\right)m^*T/B} \right]=\\
&=F(r^*)\frac{\hat{B}^\frac{5}{2}}{\sqrt{A''(r^*)}}\frac{2\pi m^*T/B}{\sinh 2\pi m^*T/B}\text{Im}\left[e^{i\pi/4}e^{i\pi \frac{A(r^*)}{\hat{B}}}e^{-i\frac{A(r^*)\pi}{2B}}\right]\sin\frac{\pi A(r^*)}{2B}=\\
&=F(r^*)\frac{\hat{B}^\frac{5}{2}}{\sqrt{A''(r^*)}}\frac{2\pi m^*T/B}{\sinh 2\pi m^*T/B}\left[\frac{\sqrt{2}}{2}-\cos\left(\frac{\pi A(r^*)}{B}-\frac{\pi}{4}\right) \right]
\end{split}
\end{equation}
The function $F(r^*)$ contains all the factors coming from the integration over the radial coordinates, like $\sqrt{g(r^*)}$ and the numerical prefactors. As can be seen, the expression found agrees with the standard result for three dimensional systems, apart from an additional and unimportant offset of the oscillations. In particular the usual scaling dimension with the magnetic field is found, i.e. $\hat{\Omega}\propto B^{5/2}$ \cite{Abrikosov} and the same frequency dependence of oscillations on the extremal section of the Fermi volume is observed.

\section{Discussions}
In a nutshell the findings of our paper is as follows:
We built, using numerical means, a charged neutron star in the $AdS_{5}$ and analyzed in detail 
the related gravitational collapse to form a black hole. We depicted the critical behavior of the 
degenerate star with few plots and then solved the Dirac equation in the spherically symmetric star 
geometry. We wrote down a flow equation governing the dynamics of fermions constructing the 
charged neutron star in $AdS_{5}$. We encountered few problems while computing the boundary 
correlation function in the presence of the star, we list the difficulties that arise in finding the 
fermionic correlator in the star background. In a sense, according to \cite{JdB:2009, JdB:2010}, it is 
an open question to find a reliable realization of the degenerate star using AdS/CFT duality and 
find a CFT interpretation of the OV limit \cite{OV:1939} in the context of collapse phenomena.  
Here we note that, if one computes the retarded Green function \ref{green} in the 
charged $AdS_{5}$ extremal RN Black Hole background then, according to \cite{Larsen:2010}, 
\begin{equation}
\label{greg}
k_{F} \approx k_{F}(0) \exp(-\sigma mL)
\end{equation}
i.e. one 
observes that the Fermi momentum fall off exponentially from the value it takes at zero fermion mass with $k_{F}(0) \simeq 0.8155$ and $\sigma \simeq 0.80$ for $AdS_{5}$ geometry,
clearly indicating a violation of Luttinger's theorem and hence proving a non-Fermi liquid behavior 
in the black hole phase. This initially prompted us to make a conclusion that one possible 
holographic interpretation one could dub to this collapse process is that when a neutron star 
collapse to form a black hole, seeing from a boundary point of view one finds a second order phase 
transition from an ideal degenerate Fermi gas to a non-Fermi liquid. Recently in a paper \cite
{Hartnoll:2010ref} the authors have found that the Kosevich-Lifshitz oscillations still persist in the charged neutron star phase.
In the same article the authors also explain how to reconcile the violation of Luttinger theorem in the case of the charged fermion star and 
the existence of the KL oscillation and the argument is still true in our $AdS_5$ case : only fermions 
in the spherical shell of radius $r^*$ contribute to the oscillations, and thus it' s not possible to 
reconstruct the whole Fermi volume from the analysis of oscillations. From the boundary field 
theory point of view this means that not all the degrees of freedom are taken into account through 
quantum oscillations, and this results in the violation of Luttinger theorem. Reproducing the 
calculations done in \cite{Hartnoll:2010ref} in our $AdS_5$ charged star, we have found again the 
appearance of the KL oscillations due to magnetic field consistent with the standard results for three dimensional systems (see \cite{Abrikosov}). After all these, it is still not clear how to realize
a holographic dual of the $AdS_5$ star, although it seems to us that both the star and the black hole are non-Fermi liquid states of matter.

Finally, despite trying to give a  concrete holographic picture of this underlying phenomena we tried 
to rethink from a different perspective, all the existing results on this subject and did a coherent 
study of all of them reaching a conclusion that this elusive phenomena still require further studies.

\section{Acknowledgements}
The work of and V.P. and R. R. has been supported in part by Dipartimento di Scienze
Fisiche of Federico II University and INFN- Section of Napoli. We would like to thank 
Finn Larsen, Greg van Anders for several correspondences during the course of the 
project. We are grateful to Sean Hartnoll for his careful reading of our manuscript and for
his valuable suggestions toward the improvement of our draft.

\begin{appendix}
\section{The zero temperature and finite chemical potential two point function in flat space}\label{ChemPot}
In our analysis above we have used the crucial fact that the effect of finite chemical potential at zero
temperature is captured by introducing a gauge field whose time component is the chemical 
potential (all other components being zero). In this section we shall try to justify this statement 
by considering fermions in flat space with a constant potential. We shall do this by considering the 
two point function of the fermions. At first we shall view, the introduction of the chemical potential,
as a redefinition of the vacuum state of theory of free fermions. Them from  there we shall 
demonstrate  that the two point function computed in this new vacuum state is the same as that 
computed for the case where the chemical potential is introduced through the time component of 
the gauge field. The later method more elegant and easy to generalize. In fact, we have used this 
above in more complicated settings where the chemical potential has a spatial variation. Therefore, in order to be certain,
\footnote{This fact is certainly true and is very well known for finite temperature. In case of finite 
temperature the chemical potential can be introduced but putting a twisted boundary condition for 
the fermion (instead of a mere anti-periodic one) along the compactified time circle. It is very well 
known that such twist can be undone by a gauge field whose time component is the chemical 
potential. This implies that the chemical potential can also introduced through such a gauge field 
without putting the twist. 
What we verify here is that it this fact continues to hold even at zero temperature.}
we present an analysis of the situation in this simple setting and verify the equivalence of the two methods.

\subsection{The free Fermi sea: The operator calculation}
In this subsection we consider free massive fermions in flat space with the fermions being filled 
upto  the fermi level with fermi momentum $k_F$. Thus the chemical potential \footnote{Here we 
shall define  the chemical potential to be the energy required to add one more particle at the fermi 
momentum} of the system is non-zero and is given by $$\mu =\sqrt{k_F^2 + m^2},$$ with $m$ 
being the mass of the fermion. 
We consider the system to be at zero temperature. We will be interested to calculate the Feynman 
propagator or the time ordered two point correlator or this system. We shall perform this through an 
operator calculation in which the two point function is the expectation value of a product of two field 
operators in a state. This state is the one in which all the single particle states upto the 
Fermi momentum is filled up. 

We consider free fermions in flat space with the system being described by the Dirac Lagrangian \cite{Peskin}
This theory is a quadratic theory and can be solved exactly. The fields $\psi$ and $\bar{\psi}$ can be expanded in terms of the creation and annihilation operators as follows
\begin{equation}
\begin{split}
 \psi(x) &= \int \frac{d^3p}{(2 \pi)^3} \frac{1}{\sqrt{2 E_p}} \sum_s \left(  a_p^s u^s(p) \exp\left({- i p.x}\right) 
         +  b_p^{s \dagger} v^s(p) \exp\left({ i p.x}\right) \right),\\
 \bar{\psi(x)} &= \int \frac{d^3p}{(2 \pi)^3} \frac{1}{\sqrt{2 E_p}} 
         \sum_s \left(  a_p^{s\dagger} \bar{u}^s(p) \exp\left({i p.x}\right) 
         +  b_p^{s} \bar{v}^s(p) \exp\left({- i p.x}\right) \right).
\end{split}
\end{equation}

Now we shall incorporate the presence of a finite chemical potential by constructing a state
in which the fermions are filled upto the energy equal to the chemical potential, characterized 
by the fermi momentum. Let us denote the state in which the fermions are filled upto the fermi 
level by $|k_F \rangle$ such that,
\begin{equation}
 |k_F \rangle = \prod_p^{k_F} \sqrt{2E_p} a^{s \dagger}_p |0 \rangle.
\end{equation}
Now for the time ordered correlation function we are required to calculate the quantities
$\langle k_F | \psi(x) \bar{\psi}(y) |k_F \rangle$ and $\langle k_F |  \bar{\psi}(y) \psi(x) |k_F \rangle$.
These quantities evaluate to
\begin{equation}
 \begin{split}
  \langle k_F | \psi(x) \bar{\psi}(y) |k_F \rangle =& \int_{k_F}^{\infty} \frac{d^3p}{(2 \pi)^3}
                      \frac{1}{2 E_p} (\displaystyle{\not{p}} + m) \exp{\left( - p.(x-y) \right)},\\
  \langle k_F |  \bar{\psi}(y) \psi(x) |k_F \rangle =& \int_{k_F}^{\infty} \frac{d^3p}{(2 \pi)^3}
                      \frac{1}{2 E_p} (\displaystyle{\not{p}} - m) \exp{\left( - p.(y-x) \right)}\\
                      &+ \int_{0}^{k_F} \frac{d^3p}{(2 \pi)^3} \left( 
                             (\displaystyle{\not{p}} + m) \exp{\left( - p.(x-y) \right)}
                     \right. \\ & \qquad \left. + (\displaystyle{\not{p}} - m) 
                                        \exp{\left( - p.(y-x) \right)} \right).
 \end{split}
\end{equation}

\begin{figure}
 \begin{center}
  \includegraphics[width=0.7\textwidth]{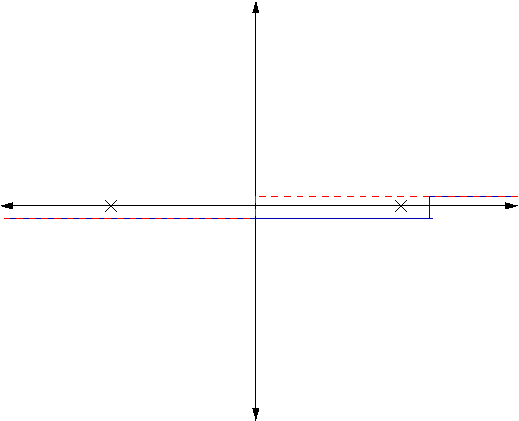}
 \end{center}
\caption{The contour prescriptions. The blue line is the new contour prescription in the presence of a finite 
constant chemical potential. The red dotted line represents the usual contour prescription.}
\label{fig:contour}
\end{figure}

Now the time-ordered two point correlation function is given by
\begin{equation}\label{twopt}
 S_F(x-y) = \left\lbrace \begin{array}{c}
            \langle k_F | \psi(x) \bar{\psi}(y) |k_F \rangle ,\ \text{for} \ x_0 > y_0,\  
						\text{(close the contour below)},\\ 
           - \langle k_F |  \bar{\psi}(y) \psi(x) |k_F \rangle ,\ \text{for} \  y_0 > x_0, \ 
						\text{(close the contour above).}
        \end{array} \right.
\end{equation}
Now the above two point function can be captured in the contour integral
\begin{equation}
 S_F(x-y) = \int \frac{d^4p}{(2 \pi)^4} \frac{i\left(\displaystyle{\not{p}}+m\right)}{p^2-m^2} \exp{-ip.(x-y)}.
\end{equation}
where the contour for $p_0$ is chosen as shown in fig:\ref{fig:contour}.
Nevertheless, the same answer can be obtained with the usual contour prescription 
(the dotted red line in fig:\ref{fig:contour})
if we include a real shift of $\mu$ in $p_0$. Therefore it is convenient to define the shifted variable,
$$ \tilde{p}_0 = p_0 - \mu$$
and make this substitution in \eqref{twopt}. Then in terms of this shifted variable $\tilde{p}_0$ the
contour prescription is the usual one.
Now in terms of this shifted variables, the effect of $\mu$ is completely captured is we introduce
a gauge field whose time component is $\mu$. This is true if we consider the eigenvalues of the 
$\partial_t$ operator to be $\tilde{p}_0$, instead of $p_0$. This justifies our use chemical potential as the 
time component of the gauge field.

\end{appendix}

\nocite{*}

\end{document}